\documentclass[a4paper,12pt]{article}
\usepackage[utf8]{inputenc}
 \usepackage[margin=1.1in]{geometry}
 \usepackage{amssymb} 
\usepackage{amsmath}
\usepackage{authblk}
\usepackage{comment}
\usepackage{tikz}
\usepackage{graphicx}
\usepackage{subfigure}
\usepackage{array}
\usepackage{cancel}
\usepackage{multirow}

\usepackage[hyperindex,breaklinks]{hyperref}
\usepackage[numbers,sort&compress]{natbib}
\usepackage{braket}
\usepackage{slashed}

\title{\Huge Fermion Dark Matter Effect on \\ Electroweak Phase Transition}
\author{\vspace*{2cm} Soudeh Mirzaie$^{1}$, Karim Ghorbani$^{1}$, Parsa Ghorbani$^{2}$}
\affil{
\small \it $^{1}$Physics Department, Faculty of Science, Arak University, Arak 38156-8-8349, Iran\\
\small\it $^{2}$Physics Department, Faculty of Science, Ferdowsi University of Mashhad, Iran\\
}

\date{}

\begin{document}
\maketitle

\begin{abstract}
The addition of extra scalars to the Standard Model (SM) of particle physics enriches the vacuum structure and consequently gives rise to strong first-order phase transitions (EWPT) in the early universe. We raise the question that how the EWPT is affected by the addition of fermions in models beyond the SM, and address this question by studying the EWPT in a dark matter model comprising a singlet scalar and two Dirac fermions.  The singlet scalar develops a nonzero vacuum expectation value (VEV), and the lighter fermion  plays the role of the dark matter. The model evades the stringent direct detection bounds due to the presence of two fermions.  
We first show that applying the high-temperature approximation, no first-order phase transition is found. Then we demonstrate  that when including the full finite temperature corrections to the effective potential, the first-order phase transition becomes possible, nevertheless, all the phase transitions will be weak. We therefore deduce that the addition of fermions reduces the strength of the EWPT. 
\end{abstract}

\newpage

\section{Introduction}
The Standard Model (SM) of particle physics suffers from different theoretical and observational inconsistencies, as such being the problem of dark matter (DM), and the matter-antimatter asymmetry in the universe, as two prominent mysteries of modern physics.
To address these problems, it seems inevitable to extend the SM by new degrees of freedom. New fields introduced in the extended models of SM, through different production mechanism (freeze-out, freeze-in, etc) accounts for the observed relic density of DM in the universe, while many direct and indirect constraints must be fulfilled. New fields also reshape the effective potential of the model and may turn the electroweak phase transition (EWPT) from crossover in the SM to a strong first-order EWPT in the extended model, as one of the Sakharov conditions (non-equilibrium) \cite{Sakharov:1967dj} for the baryogenesis. 
The simplest extension of the SM by a singlet scalar is proved to be an acceptable model of DM. The extra singlet scalar in this model augments the complexity of the vacuum structure of the effective potential which in turn results in the strong first-order EWPT \cite{Chao:2020adk,Chiang:2020yym,Ghorbani:2018yfr,Ghorbani:2020xqv,Beniwal:2018hyi,Biondini:2022ggt}. Models beyond the SM (BSM) with more scalar fields possess more complex vacuum structures and will probably always provide strong first-order EWPT. \footnote{This is not generally proved. For two-scalar  and 2HDM extension of the SM see e.g. \cite{Ghorbani:2019itr, Kanemura:2022ozv, Liu:2023sey,Ghosh:2024ing,Ghorbani:2024jaa}}

What happens to the EWPT when the SM is extended with at least one scalar and additional fermions? In this work, we address this question by studying a model of a singlet scalar and two Dirac fermion among which the lighter is the DM candidate. The EWPT and the DM in a model with an extra singlet scalar and a Dirac fermion have already been studied in \cite{Fairbairn:2013uta, Ghorbani:2017jls} demonstrating the potential for a strong first-order DM. While the addition of two Dirac fermions \cite{Ghorbani:2018hjs,Maleki:2022zuw}, helps the model evade the stringent direct detection bounds from LUX and XENONnT, we will assess the impact of these two extra fermions on the EWPT.  

In section \ref{mdl}, we introduce the two-fermion scenario and determine the physical masses in the model. In section \ref{sec_high}, we analyze the vacuum structure in the high-temperature approximation and study the various phase transition scenarios analytically. In section \ref{sec_finite}, we present the full finite-temperature effective potential.  We then review the theoretical formulation of two-fermion DM model in section \ref{2fdm}. In section \ref{rslts} we present the numerical results. Finally, we conclude in section \ref{cncl}.

\section{Model}\label{mdl}
The Lagrangian of two-fermion model at high temperature where the electroweak (EW) symmetry is unbroken reads,
\begin{equation}
\label{DM-lag}
{\cal L}_\text{DM}  = \bar \chi_1 (i {\not}\partial-\mu_1)\chi_1 + \bar \chi_2 (i {\not}\partial-\mu_2)\chi_2 
       + g_{1}~ s \bar{\chi}_{1} \chi_{1} + g_{2}~ s \bar{\chi}_{2} \chi_{2}
                          +(g_{12}~ s \bar{\chi}_{1} \chi_{2} + \text{h.c.}). 
\end{equation}
in which $\chi_1 $ and $\chi_2$ are gauge singlet Dirac fermions and $s$ is a real  singlet scalar. 
The potential part of the Lagrangian which incorporates the new singlet scalar and the SM Higgs doublet is given by,
\begin{equation}\label{tree-pot}
 V_{\text{tr}}(H,s )  =  - \frac{1}{2}\mu_{\text{s}}^{2} s^2 
                       +\frac{1}{4}\lambda_{\text{s}} s^4 + \lambda_{\text{hs}} s^2 H^{\dagger}H 
                       - \mu^{2}_{\text{h}} H^{\dagger}H + \lambda_{\text{h}} (H^{\dagger}H)^2 \,.
\end{equation}
The charges under the $\mathbb Z_2$ transformation are as follows, \footnote{The couplings of terms like $s H^\dagger H$ and $s^3$ are assumed to be negligible.}
\begin{equation}
s\to s ~, H\to H~, \chi_1 \to -\chi_1~, \chi_2 \to - \chi_2 .
\end{equation}
The bounded from below condition (BFB), is given by $\lambda_\text{h}>0,\lambda_\text{s}>0$, and 
\begin{equation}\label{bfb}
\lambda_\text{hs}>0 ~ \vee ~ (\lambda_\text{hs}<0 ~ \wedge ~ \lambda^2_\text{hs} \leq\lambda_\text{s} \lambda_\text{s} ).
\end{equation}

Both scalars take non-zero VEV, $(\braket{h}=v, \braket{s}=w)$, so there is a mixing between the Higgs and the singlet scalar 
\begin{equation}\label{ang2}
\begin{pmatrix}
h_1 \\
h_2
\end{pmatrix}=
\begin{pmatrix}
\cos\theta & \sin\theta\\
-\sin\theta & \cos\theta
\end{pmatrix}
\begin{pmatrix}
h \\
s
\end{pmatrix}.
\end{equation}
The requirement for $(v,w)$ to be the local minimum is given by
\begin{equation}
\mu_\text{h}^2=\lambda_\text{h}v^2 + \lambda_\text{hs}w^2,
~~~\mu_\text{s}^2=\lambda_\text{s}w^2 + \lambda_\text{hs}v^2.
\end{equation}
The scalars' mass matrix,
\begin{equation}
M_S^2=2 \begin{pmatrix}
 \lambda_\text{h} v^2 &    \lambda_\text{hs} vw \\
  \lambda_\text{hs} vw &  \lambda_\text{s} w^2\\
\end{pmatrix},
\end{equation}
should be diagonalized by rotating the field space by the angle $\theta$ as in Eq. (\ref{ang2}), to get the Higgs mass eigenvalue $m_1\equiv 125$ GeV, and  the singlet scalar physical mass $m_2$. 
The mixing angle in terms of the couplings and the VEVs read, 
\begin{subequations}
\begin{align}
\tan(2\theta)=\frac{2v w \lambda_\text{hs}}{\lambda_\text{s} w^2-\lambda_\text{h} v^2},
\end{align}
\end{subequations}
while the parameters $\lambda_\text{h}, \lambda_\text{s}$ and $\lambda_\text{hs}$ can be expressed in terms of the physical masses, the VEVs, and the mixing angle as 
\begin{subequations}
\begin{align}
&\lambda_\text{hs}=-\frac{(m_1^2-m_2^2)\sin(2\theta)}{2 v w },\\
&\lambda_\text{h}=\frac{(m_1^2+m_2^2)+(m_1^2-m_2^2)\cos(2\theta)}{4 v^2 },\\
&\lambda_\text{s}=\frac{(m_1^2+m_2^2)-(m_1^2-m_2^2)\cos(2\theta)}{4 w^2}.
\end{align}
\end{subequations}
In addition to scalars, there are mixing among the fermions due to the mixed Yukawa terms $s \bar\chi_1 \chi_2$ in Eq. (\ref{DM-lag}), after the scalar $s$ undergoes a non-zero VEV. The mass matrix for the fermions becomes, 
\begin{equation}
M_F= \begin{pmatrix}
-\mu_1+g_1 w&  g_{12} w  \\
g_{12} w & -\mu_2+g_2 w\\
\end{pmatrix}.
\end{equation}
Again a rotation in the fermions space as
\begin{equation}\label{angl}
\begin{pmatrix}
\psi_1 \\
\psi_2
\end{pmatrix}=
\begin{pmatrix}
\cos\xi & \sin\xi\\
-\sin\xi & \cos\xi
\end{pmatrix}
\begin{pmatrix}
\chi_1 \\
\chi_2
\end{pmatrix},
\end{equation}
leads to diagonalization of the mass matrix 
\begin{equation}
M_1=\frac{1}{2}A-\frac{1}{2}\sqrt{A^2-4B},~~~
M_2=\frac{1}{2}A+\frac{1}{2}\sqrt{A^2-4B},
\end{equation}
with 
\begin{equation}
A=(g_1+g_2)w-(\mu_1+\mu_2),~~~B=(g_1 w-\mu_1)(g_2 w-\mu_2)-g_{12}^2 w^2.
\end{equation}

\section{Electroweak Phase Transition: High-Temperature}\label{sec_high}
The potential after gauging away three components of the Higgs doublet and at very high temperature reads, 
\begin{equation}\label{tree}
 V_{\text{tr}}(h,s )  =  - \frac{1}{2}\mu_{\text{s}}^{2} s^2 
                       +\frac{1}{4}\lambda_{\text{s}} s^4 
                       - \frac{1}{2}\mu^{2}_{\text{h}} h^2 + \frac{1}{4}\lambda_{\text{h}} h^4 
                       + \frac{1}{2}\lambda_{\text{hs}} h^2 s^2 \,.
\end{equation}
The dominant one-loop thermal effective potential at high temperature approximation is, 
\begin{equation}\label{t1loop}
 V_T^{\text{1-loop}}(h,s;T)\simeq  \left( \frac{1}{2} c_\text{h} h^2 + \frac{1}{2} c_\text{s} s^2  \right) T^2 \,,
\end{equation}
where the coefficient $c_h$ and $c_s$ are,
\begin{subequations}
\begin{align}
 &c_\text{h}=\frac{1}{48}\left( 9g^2 + 3g'^2 + 12y_t^2 + 12\lambda_\text{h} + 4\lambda_{\text{hs}}  \right) \,, \\
 &c_\text{s}= \frac{1}{12}\left( \lambda_{\text{hs}} + 3\lambda_\text{s}+2g_1^2+2g_2^2+4g_{12}^2  \right)\,.
 \end{align}
\end{subequations}
The effective potential can be defined as the sum of the tree-level potential in Eq. (\ref{tree}) and the one-loop thermal contribution in Eq. (\ref{t1loop}), i.e.,  
\begin{equation}\label{eff}
 V_{\text{eff}}(h,s )  =  - \frac{1}{2}\mu_{\text{s}}^{2}(T) s^2 
                       +\frac{1}{4}\lambda_{\text{s}} s^4 
                       - \frac{1}{2}\mu^{2}_{\text{h}}(T) h^2 + \frac{1}{4}\lambda_{\text{h}} h^4 
                       + \frac{1}{2}\lambda_{\text{hs}} h^2 s^2 \,,
\end{equation}
where 
\begin{equation}
 \mu_{\text{h}}^{2}(T)=\mu_{\text{h}}^{2}-c_{\text{h}}T^2\,,~~~ \mu_{\text{s}}^{2}(T)=\mu_{\text{s}}^{2}-c_{\text{s}}T^2\,.
\end{equation}
The first derivatives of the effective potential give the extermum of the field configuration $(h,s)$. Denoting the extremum point to be $(v,w)$ the conditions are given by, 
\begin{subequations}\label{ext1}
\begin{align}
v \left( -\mu_{\text{h}}^{2}(T) +  \lambda_{\text{h}} v^2 + \lambda_{\text{hs}} w^2 \right) =0 \,, \\
w \left( -\mu_{\text{s}}^{2}(T) +  \lambda_{\text{s}} w^2 + \lambda_{\text{hs}} v^2 \right) =0.
\end{align}
\end{subequations}
There are four solutions to Eq. (\ref{ext1}) being, 
\begin{subequations}\label{sol}
 \begin{align}
 & \left( v=0\,,~ w=0 \right),\\
 & \left( v=0\,,~w^2=\frac{\mu_{\text{s}}^{2}(T)}{\lambda_{\text{s}}} \right),\\
  &\left( v^2=\frac{\mu_{\text{h}}^{2}(T)}{\lambda_{\text{h}}}\,,~ w=0 \right),\\
 & \left( v^2=  \frac{\lambda_{\text{s}} \mu_{\text{h}}^{2}(T) - \lambda_{\text{hs}} \mu_{\text{s}}^{2}(T) }{ \lambda_{\text{h}}\lambda_{\text{s}}- \lambda_{\text{hs}}^2 } \,,~w^2=\frac{\lambda_{\text{h}} \mu_{\text{s}}^{2}(T) - \lambda_{\text{hs}} \mu_{\text{h}}^{2}(T) }{ \lambda_{\text{h}}\lambda_{\text{s}}- \lambda_{\text{hs}}^2 }   \label{sol4}  \right).
 \end{align}
\end{subequations}
As the scalar field, $s$, is a mediator connecting the SM sector to the dark sector comprising two singlet fermions, it can take both zero or non-zero VEV after and before the electroweak phase transition. We therefore consider all four possible one-step phase transition scenarios being as follow, 
\begin{subequations}
 \begin{align}
 & (0,0)\to (v,0)\,,~~~(0,0)\to (v,w')\,,\\
 & (0,w)\to (v,0)\,,~~~(0,w)\to (v,w')\,,
 \end{align}
 \label{0vww'}
\end{subequations}
where $v$, $w$ and $w'$ in Eqs. (\ref{0vww'}) are non-zero VEVs. The local minimum conditions for each scenario must be held as part of the necessary criteria for having a first-order phase transition. The second derivatives of the effective potential at the extermum point are given by, 
\begin{subequations}\label{secder}
\begin{align}
 &V''_{\text{hh}}\Big\vert_{(v,w)}\equiv \frac{\partial^2 V_{\text{eff}}}{\partial^2 h} \Big\vert_{(v,w)}= 3 \lambda_{\text{h}} v^2 +\lambda_{\text{hs}} w^2 - \mu_{\text{h}}^{2}(T)\,, \\
 &V''_{\text{ss}}\Big\vert_{(v,w)}\equiv \frac{\partial^2 V_{\text{eff}}}{\partial^2 s} \Big\vert_{(v,w)}= 3 \lambda_{\text{s}} w^2 +\lambda_{\text{hs}} v^2 - \mu_{\text{s}}^{2}(T) \,,\\
 &V''_{\text{hs}}\Big\vert_{(v,w)}\equiv \frac{\partial^2 V_{\text{eff}}}{\partial h \partial s} \Big\vert_{(v,w)}= 2 \lambda_{\text{hs}} v w  \,.
 \end{align}
\end{subequations}
Now for the point $(v,w)$ in the field configuration space to be a local minimum we must have, 
\begin{equation}\label{mincon}
 V''_\text{hh}>0,~~
 \begin{vmatrix}
V''_\text{hh} &  V''_\text{hs} \\ 
 V''_\text{hs} &  V''_\text{ss} \\ 
\end{vmatrix} > 0\,.
\end{equation}
The phase transition triggers below a certain temperature at which the thermal effective potential given in Eq. (\ref{eff}) acquires two degenerate local minima, one of which stays deeper for all temperatures smaller than the critical temperature. The critical temperature then is extracted from, 
\begin{equation}\label{crt}
 V_\text{eff}(v_\text{sym}=0,w_1;T_c)=V_\text{eff}(v_\text{brk},w_2;T_c)\,.
\end{equation}

After the EWPT the VEV of the Higgs particle is non-zero and it goes down to its current value $v_\text{h}=246$ GeV at temperature near $T=0$.  In order for the phase transition to sustain continuously until $T=0$, the global minimum condition must be imposed on the minimum in the broken phase by,
\begin{equation}\label{deltaV}
 \Delta V_\text{eff}(T)\equiv V_\text{eff}(0,w_1;T)- V_\text{eff}(v_\text{brk},w_2;T)>0 \,,
\end{equation}
for $T\leqslant T_c$. This is equivalent to the condition that $T^2$-derivative of $ \Delta V_\text{eff}(T)$ at $T_c$ be positive.  Let us study each scenario in detail. 

\subsection{ Phase Transition $\left( v=0\,,~ w\neq 0 \right) \to \left( v\neq 0\,,~ w=0 \right)$ }

In this phase transition scenario the singlet scalar develops a non-zero VEV before the EWSB and a zero VEV after the EWSB.
The extrermum conditions on the thermal effective potential require that $w^2(T)= \mu_{\text{s}}^{2}(T)/\lambda_{\text{s}}$ and $v^2(T)= \mu_{\text{h}}^{2}(T)/\lambda_{\text{h}}$. Furthermore the local minimum conditions give
\begin{subequations}\label{0w}
 \begin{align}
 \mu_{\text{s}}^{2}(T)>0,  \\
    \frac{\lambda_{\text{hs}}}{\lambda_{\text{s}}} \mu_{\text{s}}^{2}(T) -\mu_{\text{h}}^{2}(T) >0,
 \end{align}
\end{subequations}
and
\begin{subequations}\label{v0}
 \begin{align}
     \mu_{\text{h}}^{2}(T)>0,  \\
    \frac{\lambda_{\text{hs}}}{\lambda_{\text{h}}} \mu_{\text{h}}^{2}(T) -\mu_{\text{s}}^{2}(T) >0\,.
 \end{align}
\end{subequations}
At the critical temperature the thermal effective potential possesses two degenerate minima. Applying the vacua in this scenario into Eq.(\ref{crt}) we have
\begin{equation}
 T_c^2=\frac{\lambda_{\text{s}}  \mu_{\text{h}}^{2} c_{\text{h}}-\lambda_{\text{h}}  \mu_{\text{s}}^{2} c_{\text{s}}-\sqrt{\lambda_{\text{h}}\lambda_{\text{s}}} |c_{\text{s}} \mu_{\text{h}}^{2}-c_{\text{h}} \mu_{\text{s}}^{2}|}{\lambda_{\text{s}}\mu_{\text{h}}^{2}-\lambda_{\text{h}} \mu_{\text{s}}^{2}}.
\end{equation}
This scenario however is not applicable here, because the scalar $s$ must get a non-zero VEV to be a SM-DM  mediator.

\subsection{ Phase Transition $\left( v=0\,,~ w=0 \right) \to \left( v\neq 0\,,~ w=0 \right)$ }\label{00v0}

This is a phase transition starting from zero VEV for both the Higgs and the scalar at high temperatures and a non-zero VEV for only the Higg field at low temperatures. In high temperature limit we have, 
\begin{subequations}\label{00}
 \begin{align}
  \mu_{\text{h}}^{2}(T)<0,\\
    \mu_{\text{s}}^{2}(T)<0, 
 \end{align}
\end{subequations}
while at low temperature the local minimum conditions become, 
\begin{subequations}\label{v0-2}
 \begin{align}
     \mu_{\text{h}}^{2}(T)>0 ,\\
    \frac{\lambda_{\text{hs}}}{\lambda_{\text{s}}} \mu_{\text{h}}^{2}(T) -\mu_{\text{s}}^{2}(T) >0\,.
 \end{align}
\end{subequations}
Eqs. (\ref{00}) and (\ref{v0}) are manifestly inconsistent, so this type of phase transition for the theory given in (\ref{tree}) does not lead to a first-order phase transition. Moreover from the dark matter point of view this scenario is also not suitable as the scalar field $s$ playing the role of the mediator between DM and SM must have a non-zero VEV after the electroweak phase transition. 

\subsection{ Phase Transition $\left( v=0\,,~ w=0 \right) \to \left( v\neq 0\,,~ w\neq 0 \right)$ } \label{00vw}

In this scenario again field configuration $(h,s)$ undergoes a change in the VEV's from a both zero before the EWPT and both non-zero after the EWPT. The non-zero VEV's after the phase transition must be the solutions given in Eq. (\ref{sol4}). The minimum condition for the point $(0,0)$ is given by Eq. (\ref{00}). The conditions for $(v,w)$ reads, 
\begin{subequations}\label{vw0}
 \begin{align}
 & 3\lambda_{\text{h}} v^2(T) +\lambda_{\text{hs}} w^2(T) - \mu_{\text{h}}^{2}(T)>0 , \\
 & \left( 3 \lambda_{\text{h}} v^2(T) +\lambda_{\text{hs}} w^2(T) - \mu_{\text{h}}^{2}(T) \right) \left( 3 \lambda_{\text{s}} w^2(T) +\lambda_{\text{hs}} v^2(T)  - \mu_{\text{s}}^{2}(T) \right)- 4 \lambda_{\text{hs}}^2 v^2(T) w^2(T)  >0,
 \end{align}
\end{subequations}
with $w$ and $v$ given by Eq. (\ref{sol4}). The conditions in Eq. (\ref{00}) and (\ref{vw0}), despite similar conditions in subsection \ref{00v0}, are not inconsistent. Therefore, we consider more necessary conditions of the first-order phase transition. Rewriting $v$ and $w$ in Eq. (\ref{sol4}) as 
\begin{equation}
 v^2(T)\equiv \alpha_1 -\beta_1 T^2\,,~~~ w^2(T)\equiv \alpha_2 -\beta_2 T^2,
\end{equation}
with 
\begin{subequations}
 \begin{align}
  \alpha_1=  \frac{\lambda_{\text{s}} \mu_{\text{h}}^{2} - \lambda_{\text{hs}} \mu_{\text{s}}^{2} }{ \lambda_{\text{h}}\lambda_{\text{s}}- \lambda_{\text{hs}}^2 }\,,~~~\beta_1= \frac{\lambda_{\text{s}} c_{\text{h}} -\lambda_{\text{hs}} c_{\text{s}} }{ \lambda_{\text{h}}\lambda_{\text{s}}- \lambda_{\text{hs}}^2 },\\
   \alpha_2=  \frac{\lambda_{\text{h}} \mu_{\text{s}}^{2} - \lambda_{\text{hs}} \mu_{\text{h}}^{2} }{ \lambda_{\text{h}}\lambda_{\text{s}}- \lambda_{\text{hs}}^2 }\,,~~~\beta_2= \frac{ \lambda_{\text{h}} c_{\text{s}} -\lambda_{\text{hs}} c_{\text{h}} }{ \lambda_{\text{h}}\lambda_{\text{s}}- \lambda_{\text{hs}}^2 },
 \end{align}
\end{subequations}
the critical temperature is as follows
\begin{equation}
 T_c^2=\frac{c_\text{h}\lambda_\text{s} \mu_\text{h}^2+c_\text{s}\lambda_\text{h} \mu_\text{s}^2 - c_\text{s}\lambda_\text{hs} \mu_\text{h}^2-c_\text{h}\lambda_\text{hs} \mu_\text{s}^2       \pm |c_\text{s} \mu_\text{h}^2-c_\text{h} \mu_\text{s}^2 | \sqrt{ \lambda_\text{hs}^2-\lambda_\text{h}  \lambda_\text{s}}}{c_\text{s}^2 \lambda_\text{h} - 2 c_\text{h} c_\text{s} \lambda_\text{hs}+c_\text{h}^2 \lambda_\text{s}},
\end{equation}
which requires $ \lambda_\text{h}  \lambda_\text{s}<\lambda_\text{hs}^2$. By looking at the BFB condition in Eq. (\ref{bfb}), this means that $\lambda_\text{hs}$ can only be positive. 

The local minimum conditions in Eqs. (\ref{00}) and (\ref{vw0}) are held for all $T\leq T_c$ if they are satisfied at $T=0$ and $T=T_c$, 
\begin{subequations}
\begin{align}
\mu^2_\text{h}<0,~~~\mu^2_\text{h}-c_\text{h}T_c^2<0,\\
\mu^2_\text{s}<0,~~~\mu^2_\text{s}-c_\text{s}T_c^2<0,
\end{align}
\end{subequations}
and 
\begin{subequations}
\begin{align}
& 3\lambda_{\text{h}} \alpha_1 +\lambda_{\text{hs}} \alpha_2 - \mu_{\text{h}}^{2}>0,  \\
 & \left( 3 \lambda_{\text{h}} \alpha_1 +\lambda_{\text{hs}} \alpha_2 - \mu_{\text{h}}^{2}\right) \left( 3 \lambda_{\text{s}} \alpha_2 +\lambda_{\text{hs}} \alpha_1 - \mu_{\text{s}}^{2} \right)- 4 \lambda_{\text{hs}}^2 \alpha_1\alpha_2 >0,\\
& 3\lambda_{\text{h}} \alpha_1 +\lambda_{\text{hs}} \alpha_2 - \mu_\text{h}^{2}+(c_\text{h}-3\lambda_\text{h} \beta_1-\lambda_\text{hs}\beta_2)T_c^2>0 , \\
\begin{split}
& \left( 3\lambda_{\text{h}} \alpha_1 +\lambda_{\text{hs}} \alpha_2 - \mu_\text{h}^{2}+(c_\text{h}-3\lambda_\text{h} \beta_1-\lambda_\text{hs}\beta_2)T_c^2 \right) \\
&\times \left( 3\lambda_{\text{s}} \alpha_2 +\lambda_{\text{hs}} \alpha_1 - \mu_\text{s}^{2}+(c_\text{s}-3\lambda_\text{s} \beta_2-\lambda_\text{hs}\beta_1)T_c^2 \right)- 4 \lambda_{\text{hs}}^2 (\alpha_1 -\beta_1 T_c^2) (\alpha_2 -\beta_2 T_c^2)  >0.
\end{split}
\end{align}
\end{subequations}

The global minimum condition for the point $(v,w)$ reads
\begin{equation}
\begin{split}
 & c_\text{h}\lambda_\text{s} \mu^2_\text{h}+c_\text{s}\lambda_\text{h} \mu^2_\text{s} -c_\text{s} \lambda_\text{hs} \mu^2_\text{h}-c_\text{h} \lambda_\text{hs} \mu^2_\text{s}+\left( 2 c_\text{h} c_\text{s}  \lambda_\text{hs} -c^2_\text{s} \lambda_\text{h}-c^2_\text{h}\lambda_\text{s}  \right)T_c^2>0
   \end{split}.
\end{equation}

\subsection{ Phase Transition $\left( v=0\,,~ w\neq 0 \right) \to \left( v\neq 0\,,~ w'\neq 0 \right)$ }

The phase transition in this scenario is from the minimum $(0,w)$ with $w^2=\mu_{\text{s}}^{2}(T)/\lambda_{\text{s}}$ at high temperature into the minimum $(v,w')$ with  
\begin{equation}
v^2=  \frac{\lambda_{\text{s}} \mu_{\text{h}}^{2}(T) - \lambda_{\text{hs}} \mu_{\text{s}}^{2}(T) }{ \lambda_{\text{h}}\lambda_{\text{s}}- \lambda_{\text{hs}}^2 },~~~~~ 
w'^2=\frac{\lambda_{\text{h}} \mu_{\text{s}}^{2}(T) - \lambda_{\text{hs}} \mu_{\text{h}}^{2}(T) }{ \lambda_{\text{h}}\lambda_{\text{s}}- \lambda_{\text{hs}}^2 } 
\end{equation}
at low temperature. The first-order conditions for the point $(0,w)$ reads,
\begin{subequations}\label{0w-2}
 \begin{align}
   \frac{\lambda_{\text{hs}}}{\lambda_{\text{s}}} \mu_{\text{s}}^{2}(T) -\mu_{\text{h}}^{2}(T) >0 \,,\\
   \mu_{\text{s}}^{2}(T)>0\,.
 \end{align}
\end{subequations}
The same conditions for the minimum $(v,w')$ becomes, 
\begin{subequations}\label{vw'}
 \begin{align}
\frac{ \lambda_{\text{h}} \left( \lambda_{\text{s}} \mu_{\text{h}}^{2}(T) - \lambda_{\text{hs}} \mu_{\text{s}}^{2}(T) \right)}{ \lambda_{\text{h}}\lambda_{\text{s}}- \lambda_{\text{hs}}^2 } >0 \,,\\
  \frac{ \left( \lambda_{\text{h}} \mu_{\text{s}}^{2}(T) - \lambda_{\text{hs}} \mu_{\text{h}}^{2}(T) \right) \left( \lambda_{\text{s}} \mu_{\text{h}}^{2}(T) - \lambda_{\text{hs}} \mu_{\text{s}}^{2}(T) \right)}{ \lambda_{\text{h}}\lambda_{\text{s}}- \lambda_{\text{hs}}^2 } > 0\,.
 \end{align}
\end{subequations}
The critical temperature for this scenarios is given by 
\begin{equation}
T_c^2=\frac{ a \pm \sqrt{ a^2+bc}}{b}
\end{equation}
where 
\begin{subequations}
 \begin{align}
 & a=c_{\text{h}} \alpha_1 \lambda_\text{s}  + 
 c_\text{s}( \alpha_2 \lambda_\text{s}- \mu^2_\text{s}) - \lambda_\text{s}( \alpha_1 \beta_1 \lambda_\text{h} +
\alpha_2 \beta_1 \lambda_\text{hs} 
+ \alpha_1 \beta_2 
\lambda_\text{hs} + \alpha_2 \beta_2 \lambda_\text{s}- \beta_1 \mu^2_\text{h} - \beta_2 \mu^2_\text{s}),\\
&  b=-c^2_\text{s}+\lambda_\text{s}(2c_\text{h} \beta_1 +
 2c_\text{s} \beta_2 -\lambda_\text{h} \beta_1^2 -
    2\lambda_\text{hs} \beta_1 \beta_2 - \beta^2_2),   \\
& c= \lambda_\text{s} \alpha_1  ( \lambda_\text{h}  \alpha_1+ 
 2 \lambda_\text{hs}  \alpha_2- 2\mu_\text{h}^2) +(\mu_\text{s}^2-\lambda_\text{s} \alpha_2)^2.
 \end{align}
\end{subequations}
Both Eqs. (\ref{0w-2}) and (\ref{vw'}) must be satisfied for all $T\leqslant T_c$, which means they must be satisfied for $T=0$ and $T=T_c$, i.e. 
\begin{subequations}
\begin{align}
&\frac{\lambda_{\text{hs}}}{\lambda_{\text{s}}} \mu_{\text{s}}^{2} -\mu_{\text{h}}^{2} >0, \\
&\mu_{\text{s}}^{2}>0,\\
&\frac{\lambda_{\text{hs}}}{\lambda_{\text{s}}} (\mu_{\text{s}}^{2}-c_\text{s}T_c^2) -(\mu_{\text{h}}^{2}-c_\text{h}T_c^2) >0, \\
&\mu_{\text{s}}^{2}-c_\text{s}T_c^2>0,
\end{align}
\end{subequations}
and 
\begin{subequations}
 \begin{align}
\frac{ \lambda_{\text{h}} \left( \lambda_{\text{s}} \mu_{\text{h}}^{2} - \lambda_{\text{hs}} \mu_{\text{s}}^{2} \right)}{ \lambda_{\text{h}}\lambda_{\text{s}}- \lambda_{\text{hs}}^2 } >0 \,,\\
  \frac{ \left( \lambda_{\text{h}} \mu_{\text{s}}^{2} - \lambda_{\text{hs}} \mu_{\text{h}}^{2} \right) \left( \lambda_{\text{s}} \mu_{\text{h}}^{2} - \lambda_{\text{hs}} \mu_{\text{s}}^{2} \right)}{ \lambda_{\text{h}}\lambda_{\text{s}}- \lambda_{\text{hs}}^2 } > 0\,.
 \end{align}
\end{subequations}
Now, the global minimum condition takes the following form, 
\begin{equation}
\begin{split}
 & c_\text{s} \alpha_2 \lambda_\text{s}  - c_\text{s} \mu_\text{s}^2 + c_\text{h} \alpha_1 \lambda_\text{s} - \alpha_2 \beta_1 \lambda_\text{hs} - \alpha_1 \beta_1 \lambda_\text{h} - \alpha_1 \beta_2 \lambda_\text{hs} -\alpha_2 \beta_2 \lambda_\text{s} +\beta_1 \mu_\text{h}^2+\beta_2  \mu_\text{s}^2 \\
 & + \left( c_\text{s}^2-2 c_\text{s} \beta_2 \lambda_\text{s}-2 \beta_1 \lambda_\text{s} +\beta_1^2 \lambda_\text{h}+2\beta_1 \beta_2 \lambda_{hs} + \beta_2^2 \lambda_\text{s}
   \right)T_c^2>0.
   \end{split}
\end{equation}

\section{Electroweak Phase Transition: Finite-Temperature}\label{sec_finite}

The finite temperature effective potential beside the tree-level potential consists of thermal contributions and zero-temperature Coleman-Weinberg one-loop correction,
\begin{equation}
V_\text{eff}(h,s;T)=V_\text{tr}(h,s)+V_\text{CW}(h,s)+V_T(h,s;T).
\end{equation}
For the two-fermion DM model with an extra singlet scalar and two singlet Dirac fermions, the one-loop Coleman-Weinberg contribution $V_\text{CW}(h,s)$ using the on-shell renormalization scheme and cut-off regularization reads \cite{Coleman:1973jx} (see also \cite{Quiros:1999jp}),
\begin{equation}\label{Vcw}
V_\text{CW}=\frac{1}{64 \pi^2}\sum_i{(-1)^{f_i} g_i \left[ M^4_i(h,s) \left(\log\frac{M^2_i(h,s)}{M^2_i(v,w)} -\frac{3}{2}\right) - 2M^2_i(h,s)  M^2_i(v,w)\right]},
\end{equation}
where $M_i(h,s)$ are scalar-field-dependent masses of all particles in the  model
\begin{subequations}
\label{fieldmass}
\begin{align}
&M^2_\text{h,s}=\begin{pmatrix}
3 \lambda_{\text{h}} h^2 +\lambda_{\text{hs}} s^2 - \mu_{\text{h}}^{2} &  2 \lambda_{\text{hs}} h s , \\
2 \lambda_{\text{hs}} h s  & 3 \lambda_{\text{s}} s^2 +\lambda_{\text{hs}} h^2 - \mu_{\text{s}}^{2}
\end{pmatrix},\\
&M^2_{\psi_1,\psi_2}= \begin{pmatrix}
-\mu_1+g_1 s&  g_{12} s  \\
g_{12} s & -\mu_2+g_2 s\\
\end{pmatrix},\\
&M^2_W=\frac{1}{4}g^2 h^2,\\
&M^2_Z=\frac{1}{4}(g^2+g'^2)h^2,\\
&M^2_t=\frac{1}{2}y^2_t h^2. 
\end{align}
\end{subequations} 
where $i$ stands for the $i$-th particle in Eq. (\ref{Vcw}). For the mass matrices in Eqs. (\ref{fieldmass}), only the relevant mass eigenvalues should be inserted in Eq. (\ref{Vcw}).  For bosons $f_i=0$ and for fermions $f_i=1$. Also $M_i(v,w)$ stands for $i$-th particle mass evaluated at the Higgs and the singlet scalar VEV. The thermal one-loop potential $V_T$ is given by \cite{Dolan:1973qd, Quiros:1999jp},
\begin{equation}
V_T=\frac{T^4}{2\pi^2}\sum_i{(-1)^{f_i}g_i J_{F,B}\left(\frac{M_i}{T}\right)},
\end{equation}
where
\begin{equation}
J_{F,B}(x)=\int_0^\infty{y^2\log{\left(1\pm e^{-\sqrt{x^2+y^2}}\right)dy }},
\end{equation}
where $M_i$ is the field dependent mass, and $B$ ($F$) stands for boson (fermion) and corresponds to the minus (plus) sign. 
In order to include the contributions of daisy diagrams to prevent the IR divergence at high temperature we need to modify the total effective potential by replacing the masses in $V_\text{CW}$ and $V_T$ by thermally-corrected masses as
\begin{subequations}
\begin{align}
&M^2_\text{h,s}\to M^2_\text{h,s}+\begin{pmatrix}
\Pi_\text{h}(T) & 0 \\
0 & \Pi_\text{s}(T)
\end{pmatrix},\\
&M^2_W(h)\to M^2_W(h) + \Pi_W(T),\\
&M^2_{Z,\gamma}(h)=\begin{pmatrix}
\frac{1}{4}g^2 h^2+\frac{11}{6}g^2 T^2 & -\frac{1}{4} g g' h^2 \\
-\frac{1}{4} g g' h^2 & \frac{1}{4}g'^2 h^2+\frac{11}{6}g'^2 T^2
\end{pmatrix},
\end{align}
\end{subequations}
with the thermal mass contributions $\Pi_i(T)$
\begin{subequations}
\begin{align}
&\Pi_\text{h}(T)=\frac{1}{48}\left(9g^2+3g'^2+12y_t^2+24\lambda_\text{h}+2\lambda_\text{hs} \right) T^2,\\
&\Pi_\text{s}(T)=\frac{1}{12}\left(3\lambda_\text{s}+2\lambda_\text{hs} \right)T^2,\\
&\Pi_W(T)=\frac{11}{6}g^2 T^2.
\end{align}
\end{subequations}
\begin{figure}
\centering
\includegraphics[scale=.5, angle=-90]{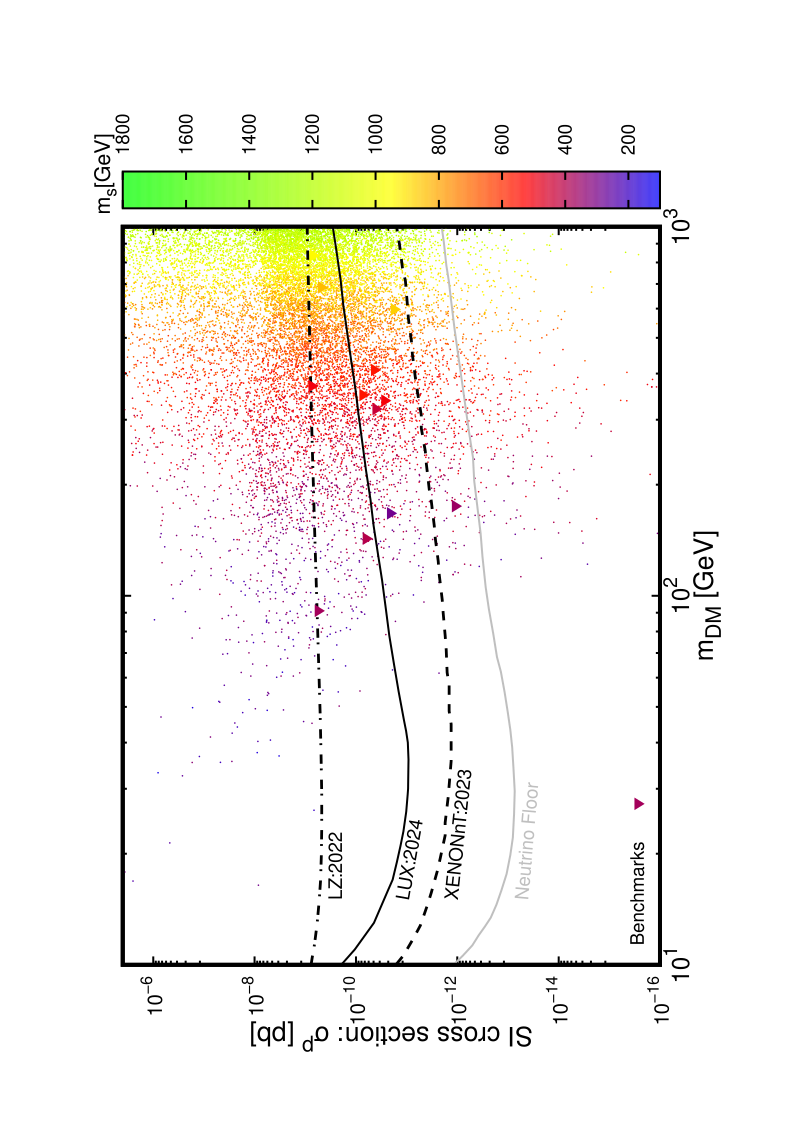}
\caption{Shown is the DM mass vs. DM-nucleon cross section when the DM relic density is constrained to $\Omega_\text{DM}h^2 \sim 0.12$. The singlet scalar mass is shown as color spectrum. The DM-nucleon direct detection bounds XENONnT,  LZ and LUX, exclude part of the parameter space. The benchmarks are the points in Table 2, corresponding to first-order electroweak phase transition. }
\label{dm_nuc}
\end{figure}
\section{Two-Fermion Dark Matter}\label{2fdm}
In the two-fermion model after diagonalization of the fermion mass matrix by rotating the fermions space from $(\chi_1,\chi_2)$ to $(\psi_1,\psi_2)$ with physical mass splitting $\delta=M_2-M_1$ the lighter fermion, $\psi_1$, is taken as the DM candidate. The singlet scalar  $h_2$, plays the role of the mediator between the dark and visible sectors. This model is interestingly able to evade the direct detection strong bounds \cite{Ghorbani:2018hjs}.
\subsection{Dark Matter Abundance}
\begin{figure}[htbp]
    \centering
    \subfigure[$w=34$ GeV]{\includegraphics[width=0.3\textwidth]{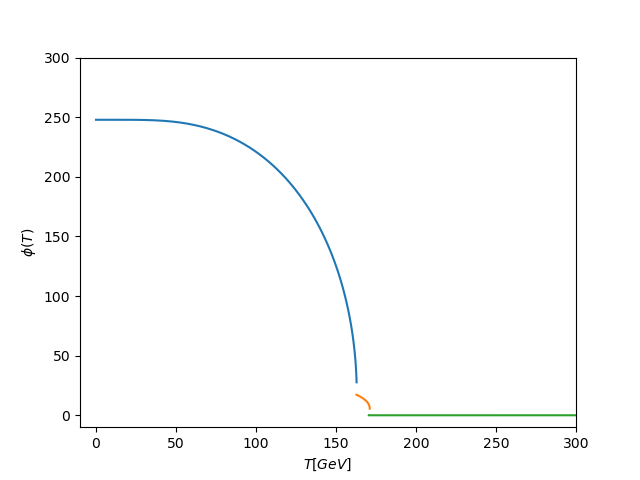}}
    \subfigure[$w=40$ GeV]{\includegraphics[width=0.3\textwidth]{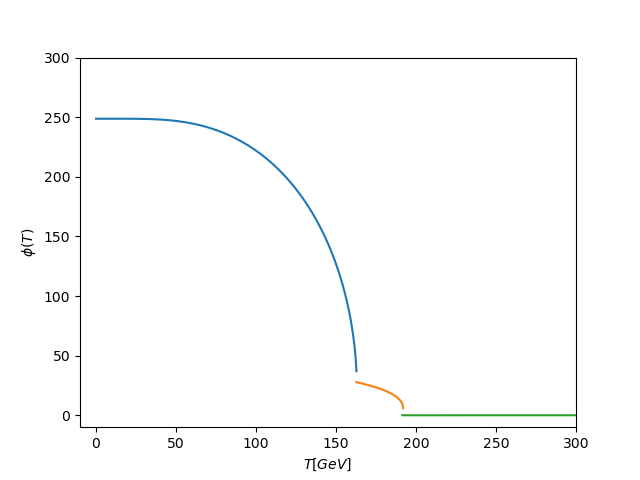}}
    \subfigure[$w=53$ GeV]{\includegraphics[width=0.3\textwidth]{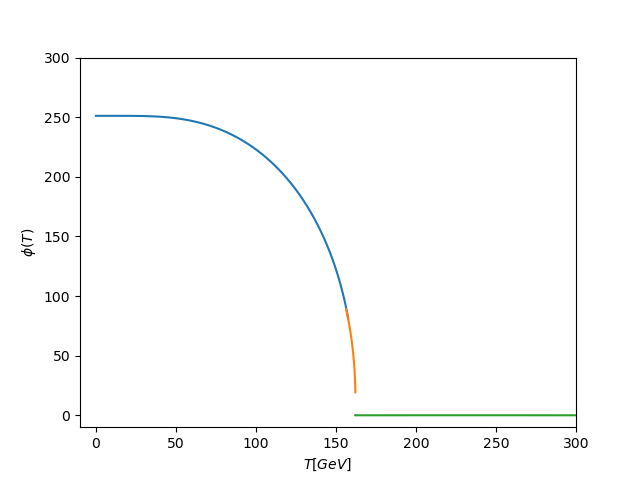}}
    \subfigure[$w=60$ GeV]{\includegraphics[width=0.3\textwidth]{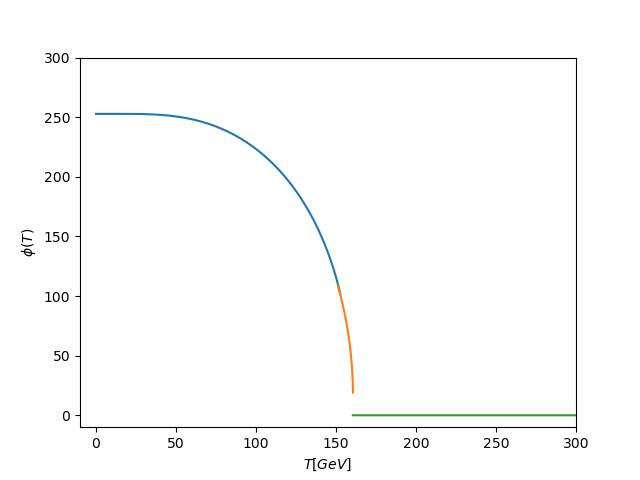}}
    \subfigure[$w=70$ GeV]{\includegraphics[width=0.3\textwidth]{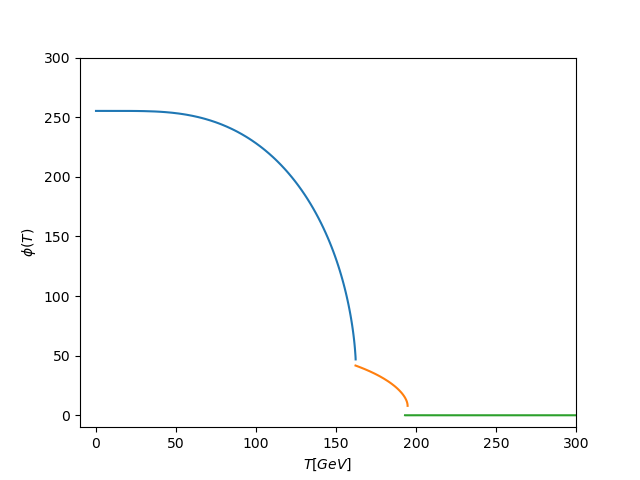}}
    \subfigure[$w=85$ GeV]{\includegraphics[width=0.3\textwidth]{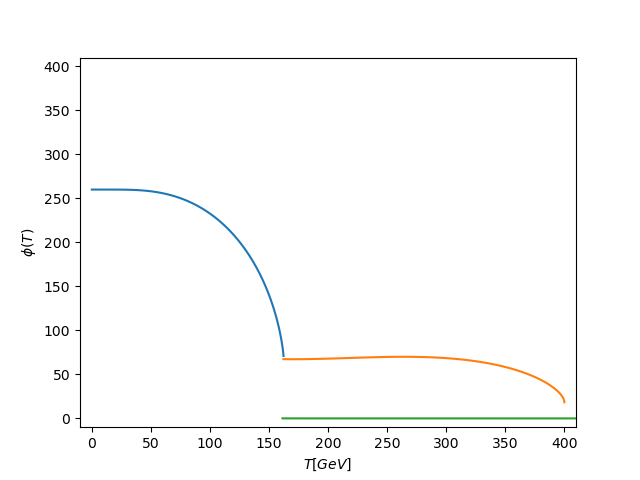}}
    \subfigure[$w=93$ GeV]{\includegraphics[width=0.3\textwidth]{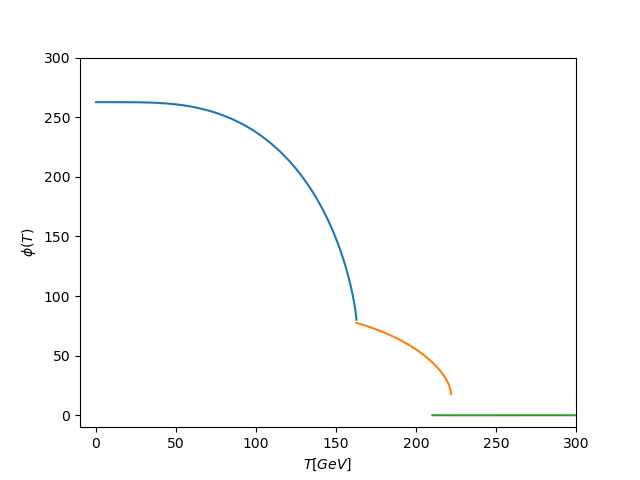}}
    \subfigure[$w=103$ GeV]{\includegraphics[width=0.3\textwidth]{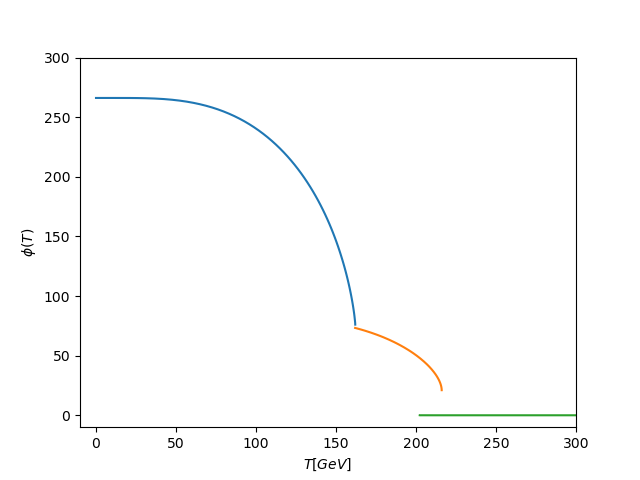}}
    \subfigure[$w=112$ GeV]{\includegraphics[width=0.3\textwidth]{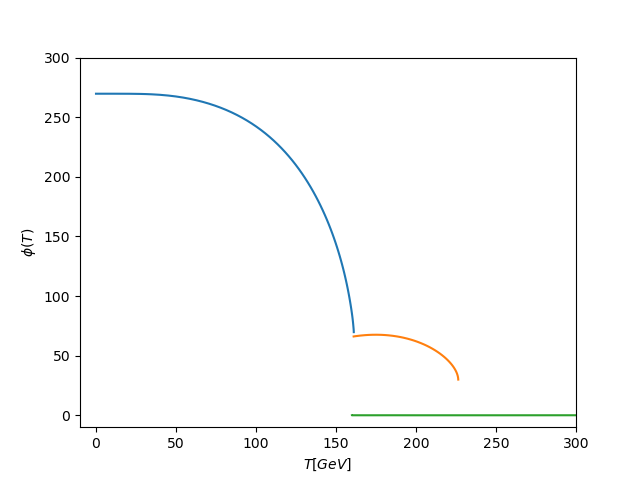}}
     \subfigure[$w=122$ GeV]{\includegraphics[width=0.3\textwidth]{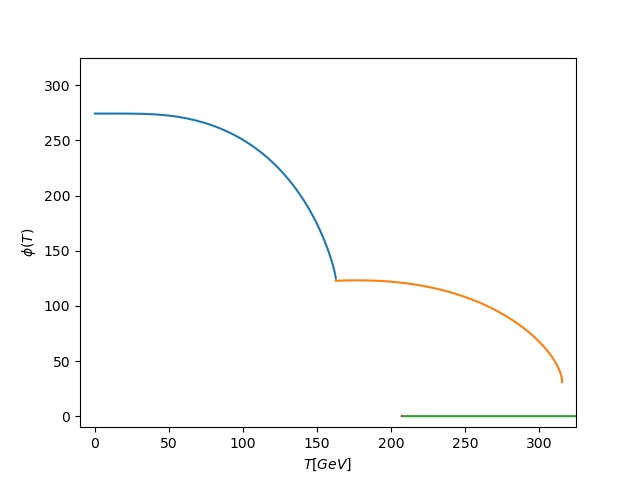}}
    \subfigure[$w=132$ GeV]{\includegraphics[width=0.3\textwidth]{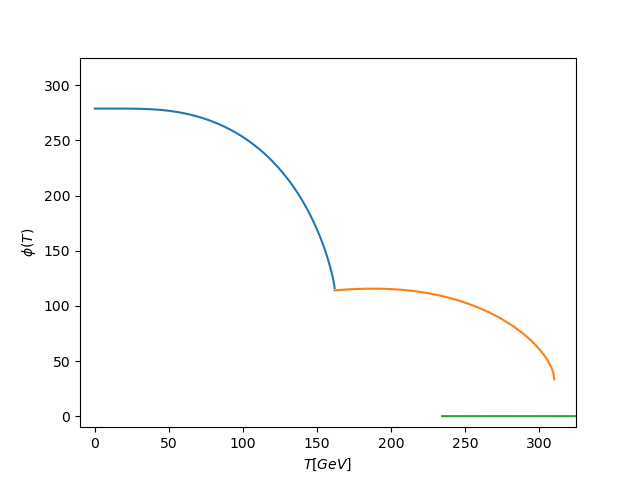}}
    \subfigure[$w=142$ GeV]{\includegraphics[width=0.3\textwidth]{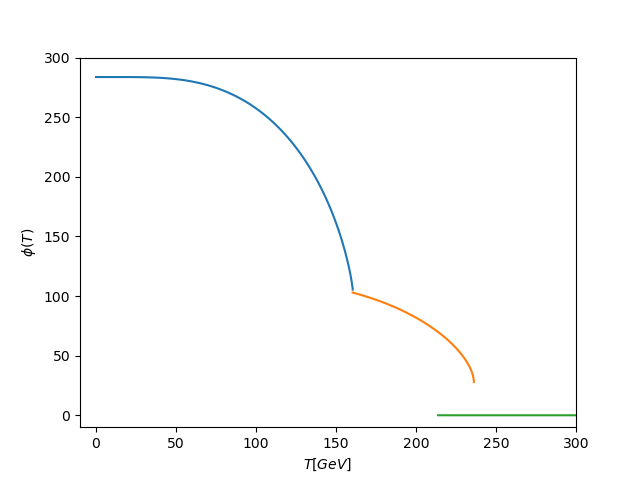}}
    \caption{Shown are two-step first-order electroweak phase transitions in two-fermion dark matter model for different scalar VEVs. All the phase transitions are weak.}
    \label{2stepphase}
\end{figure}
The DM fermion annihilation into SM fermions occurs via $s$-channel and into the Higgs and the singlet scalar goes through the $t$- and $u$-channels. Moreover, due to the fermion mixing term in the model, there is also coannihilation into the SM fermions and the scalars. To take into account the effect of the coannihilation in the DM number density the Boltzmann equation can be solved with an effective coannihilation cross section \cite{PhysRevD.43.3191,Edsjo:1997bg}  
\begin{equation}
\frac{dn}{dt}+3nH=-\braket{\sigma_\text{eff}v}\left(n^2-n_\text{eq}^2 \right),
\end{equation}
where $n=n_1+n_2$ is total number density incorporating the DM particle $\psi_1$ and the heaver fermion $\psi_2$. Also $H$ stands for the Hubble parameter and $\sigma_\text{eff}$ is the effective DM annihilation cross  section given by 
\begin{equation}
\sigma_\text{eff}=\frac{4}{g_\text{eff}^2}\left( \sigma_{11}+\sigma_{22} \left(1+\frac{\delta}{M_1} \right)^{3} e^{-2\delta/T}+2\sigma_{12}\left(1+\frac{\delta}{M_1} \right)^{3/2} e^{-\delta/T} \right),
\end{equation}
with $\sigma_{ij}$ being the (co)annihilation cross section for $\psi_i \psi_j \to SM$, and $g_\text{eff}$ being the effective internal degrees of freedom given by 
\begin{equation}
g_\text{eff}=2+2\left( 1+\frac{\delta}{M_1}\right)^{3/2}e^{-\delta/T}.
\end{equation}
The bracket $\braket{}$ means the thermal average and $\delta$ is the mass difference $\delta=M_2 - M_1$.   
To solve the above Boltzmann equation we use the {\tt MicrOMEGAs} package. The free parameters in the model are the masses $m_2, M_1, M_2$, the couplings $g_1, g_2, g_{12}$, and the mixing angle $\theta$ and the scalar VEV, $w$. The Higgs mass and the Higgs VEV take their experimental values  $m_1=125$ GeV and $v_0=246$ GeV. 
\subsection{Direct Detection}
\begin{table}[t]
\renewcommand{\familydefault}{\rmdefault}
\renewcommand{\arraystretch}{2} 
\setlength{\tabcolsep}{2pt}     
\centering
\resizebox{\textwidth}{!}{                          
\begin{tabular}{|c|c|c|c|c|c|c|c|c|c|}
\hline
      & \multicolumn{4}{c|}{Phase Transition -- Step 1} & \multicolumn{5}{c|}{Phase Transition -- Step 2} 
\\ \hline
\ $\mathit{w}$(GeV) & $\mathit{T_n}$(GeV) & $\mathit{T_c}$(GeV) & From: $(v, w)$ &  To: $(v, w)$  & $\mathit{T_n}$(GeV) & $\mathit{T_c}$(GeV) &  From: $(v, v)$  &  To: $(v, w)$ & $v(T_n)/T_n$ \\ \hline
$\mathbf{33.87}$  & $170.98$  & $170.99$ & $(0 , 0)$ & $(0 , 7.05)$  & $162.83$  & $162.84$ & $(0 , 17.10)$ &$(23.79 , 17.11)$ & $0.146$ \\ \hline
$\mathbf{40.00}$  & $191.86$  & $191.87$ & $(0 , 0)$  & $(0 ,7.51)$  & $162.76$  & $162.77$ & $(0 , 27.70)$ & $(24.44 , 27.7)$ & $0.150$ \\ \hline
$\mathbf{53.09}$  & $162.04$  & $162.05$  & $(0 , 0)$  & $(23 , 0)$  &  $157.36$ & $157.38$ & $(82 , 0)$ & $(82.58 , 8.52)$ & $0.145$ \\ \hline
$\mathbf{60.57}$  & $160.56$  & $160.57$ & $(0 , 0)$  & $(23 , 0)$ & $152.3$  & $152.4$  & $(100 , 0)$ & $(102.8 , 12.3)$ & $0.146$ \\ \hline
$\mathbf{70.07}$ & $194.47$  & $194.5$ & $(0 , 0)$  & $(0 , 10)$  & $162.24$  & $162.25$ & $(0 , 41)$ & $(23.71 , 41.81)$ & $0.146$ \\ \hline
$\mathbf{85.90}$  & $396.3$  & $396.7$ & $(0 , 0)$  & $(0 , 28)$ &  $162.29$ & $162.3$ & $(0 , 67)$ & $(23.71 , 67.40)$ & $0.146$ \\ \hline
$\mathbf{93.34}$  & $220.4$  & $220.6$ & $(0 , 0)$ & $(0 , 25)$  & $162.71$  & $162.72$ & $(0 , 77)$ & $(23.6 , 77.5)$ & $0.146$ \\ \hline
$\mathbf{103.0}$  & $214.1$  & $214.4$ & $(0 , 0)$  & $(0 , 29)$  &  $161.98$  & $161.99$ & $(0 , 73)$ & $(23.7 , 73.5)$ & $0.146$ \\ \hline
$\mathbf{112.1}$ &  $222.0$  & $222.9$  & $(0 , 0)$ & $(0 , 42)$ & $161.07$  & $161.08$  & $(0 , 66)$ & $(24 , 66.9)$  & $0.149$ \\ \hline
$\mathbf{122.40}$ & $311.4$ & $312.1$ & $(0 , 0)$ & $(0 , 48)$ & $162.77$ & $162.78$ & $(0 , 120)$ & $(23.7 , 122.7)$ & $0.149$ \\ \hline
$\mathbf{132.30}$ & $305.6$ & $306.4$ & $(0 , 0)$ & $(0 , 50)$ & $161.89$ & $161.90$& $(0 , 110)$ & $(23.9 , 114.2)$ & $0.145$ \\ \hline
$\mathbf{142.40}$ & $233.5$ & $233.9$ & $(0 , 0)$ & $(0 , 40)$ & $160.411$ & $160.419$& $(0 , 100)$ & $(23.8 , 103.2)$ & $0.147$ \\ \hline
\end{tabular}
}
\caption{The vacuum structure, the critical and nucleation temperature of  two-step electroweak phase transitions for different values of the scalar VEV corresponding to phase transition diagrams in Figure \ref{2stepphase} are shown.}
\label{phasepar1}
\end{table}
\begin{table}[t]
\renewcommand{\familydefault}{\rmdefault}
\centering
\setlength{\tabcolsep}{1pt} 
\renewcommand{\arraystretch}{1.2} 
\begin{tabular}{|c|c|c|c|c|c|c|c|c|c|c|}
\hline
$w$(GeV) & $M_1$(GeV) &  $M_2$(GeV)  &  $m_s$(GeV) & 
$\lambda_h$ & $\lambda_{hs}$ & $\lambda_s$ & $g_1$ & $g_2$ & $g_{12}$ & $\sin \theta$ \\
\hline
$\mathbf{33.87}$ & $409.10$ & $561.40$ & $144.9$ & $0.1291$ & $0.0214$ & $9.152$ & $1.244$ & $1.994$ & -$0.208$ & $0.03$ \\ \hline
$\mathbf{40.00}$ & $596.7$ & $871.6$ & $164.2$ & $0.1291$ & $0.0172$ & $8.42$ & $1.072$ & $0.9554$ & -$0.873$ & $0.01$ \\ \hline
$\mathbf{53.09}$ & $836.20$ & $943.50$ & $86.59$ & $0.1290$ & -$0.0212$ & $1.332$ & $1.154$ & $0.348$ & $0.799$ & $0.03$ \\ \hline
$\mathbf{60.57}$ & $349.80$ & $558.30$ & $85.22$ & $0.1295$ & -$0.0516$ & $0.993$ & $0.130$ & $1.487$ & $0.971$ & $0.09$ \\ \hline
$\mathbf{70.07}$ & $167.30$ & $287.80$ & $95.62$ & $0.1290$ & -$0.0153$ & $0.932$ & $0.254$ & $1.191$ & $0.690$ & $0.05$ \\ \hline
$\mathbf{85.90}$ & $685.00$ & $838.20$ & $85.90$ & $0.1290$ & -$0.0153$ & $0.532$ & $0.856$ & $1.180$ & $0.953$ & $0.04$ \\ \hline
$\mathbf{93.34}$ & $175.00$ & $358.90$ & $67.55$ & $0.1291$ & -$0.0034$ & $0.262$ & $0.100$ & $0.243$ & $0.749$ & $0.006$ \\ \hline
$\mathbf{103.00}$ & $142.80$ & $400.20$ & $67.50$ & $0.1290$ & -$0.0156$ & $0.215$ & $0.147$ & $1.877$ & $0.781$ & $0.03$ \\ \hline
$\mathbf{112.10}$ & $320.70$ & $435.20$ & $48.19$ & $0.1289$ & -$0.0218$ & $0.093$ & $0.039$ & $1.325$ & -$0.883$ & $0.04$ \\ \hline
$\mathbf{122.40}$ & $337.90$ & $508.60$ & $75.37$ & $0.1291$ & -$0.0053$ & $0.189$ & $0.299$ & $0.904$ & $0.886$ & $0.01$ \\ \hline
$\mathbf{132.30}$ & $369.90$ & $502.40$ & $74.76$ & $0.1288$ & -$0.0181$ & $0.160$ & $0.414$ & $1.122$ & $0.864$ & $0.06$ \\  \hline
$\mathbf{142.40}$ & $91.02$ & $373.40$ & $86.40$ & $0.1285$ & -$0.2257$ & $0.1859$ & $0.3588$ & $1.052$ & $0.689$ & $0.09$ \\ \hline
\end{tabular}
\caption{The values of the free parameters in the model corresponding to the two-step phase transitions shown in Figure \ref{2stepphase}.}
\label{phasepar2}
\end{table}
The direct detection (DD) of the DM through DM scattering off the nucleus has become more stringent than ever in the recent DD experiments \cite{LUX:2016ggv, PandaX-4T:2021bab,LZ:2022lsv,XENON:2023cxc}, which brings the lower limits of the DM-nucleon cross section almost down to the neutrino floor \cite{Billard:2021uyg}. Therefore many DM models including the singlet fermion DM are already ruled out. However, similar to the case of two-scalar scenario, the two-fermion model evades the strong DD limits as explained in \cite{Ghorbani:2018hjs}.  
The DM-nucleon cross section is deduced from a $t$-channel interaction among the DM fermion and quarks inside the nucleon $\psi_1 q \to \psi_1 q$, mediated by the Higgs and the singlet scalar
\begin{equation}
\mathcal{L}= c_q \bar\psi_1 \psi_1 \bar q q,
\end{equation}
where the coupling is given by
\begin{equation}
c_q=g_1 \sin(2\theta)\frac{m_q}{2v_0}\left( \frac{1}{m^2_\text{h}}-\frac{1}{m^2_\text{s}} \right).
\end{equation}
The spin-independent (SI) DM-nucleon scattering cross section at zero momentum transfer is given
\begin{equation}
\sigma^N_\text{SI}=\frac{4\mu_N^2 c_N^2}{\pi},
\end{equation}
where $\mu_N$ is the reduced mass of the  DM the nucleon and the nucleon effective coupling $c_N$, is expressed in terms of the quark coupling $c_q$ and scalar form factors  \cite{Ellis:2008hf,Crivellin:2013ipa},
\begin{equation}
c_N=m_N\left( \sum_{q=u,d,s}f_{Tq}^N \frac{c_q}{m_q}+\frac{2}{27}f_{Tg}^N\sum_{q=c,b,t}\frac{c_q}{m_q} \right),
\end{equation}
with scalar form factors being 
\begin{equation}
 f_u^p=0.0153,~ ~ f_d^p=0.0191,~~ f_s^p=0.0447.
\end{equation}
The numerical calculations again is done using the package {\tt MicrOMEGAs}.
We confront the model against a combination of the constraints of the DD bounds, DM relic abundance, and the conditions for a strongly first-order phase transition. 
\begin{figure}[t]
    \centering
    \subfigure[$w=128$ GeV]{\includegraphics[width=0.3\textwidth]{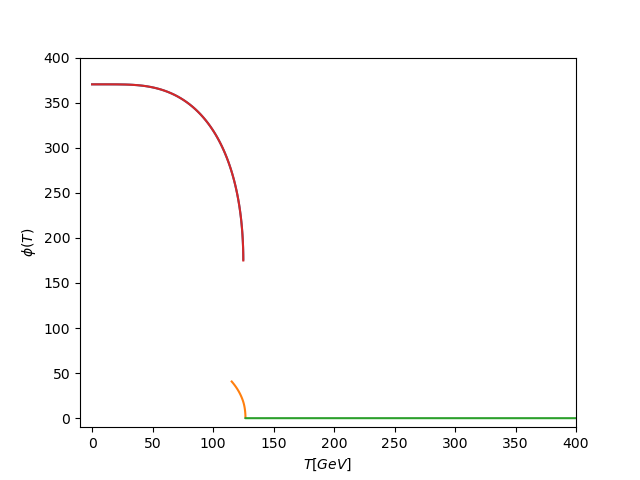}}
    \subfigure[$w=198$ GeV]{\includegraphics[width=0.3\textwidth]{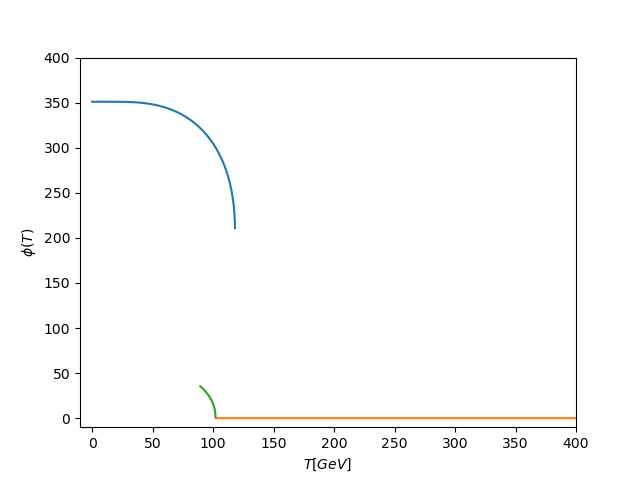}}
    \subfigure[$w=248$ GeV]{\includegraphics[width=0.3\textwidth]{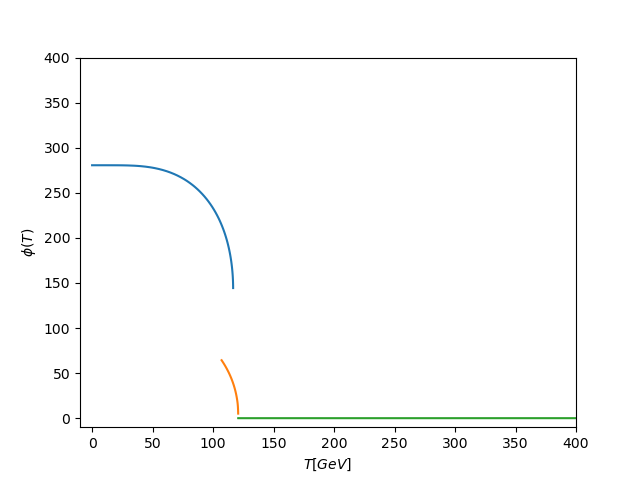}}
    \subfigure[$w=254$ GeV]{\includegraphics[width=0.3\textwidth]{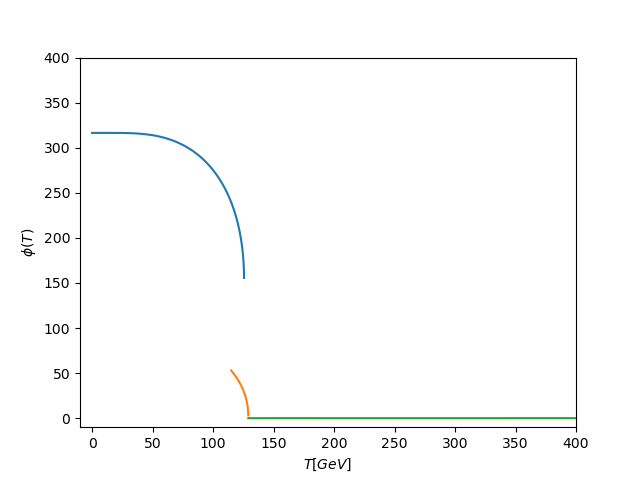}}
    \subfigure[$w=259$ GeV]{\includegraphics[width=0.3\textwidth]{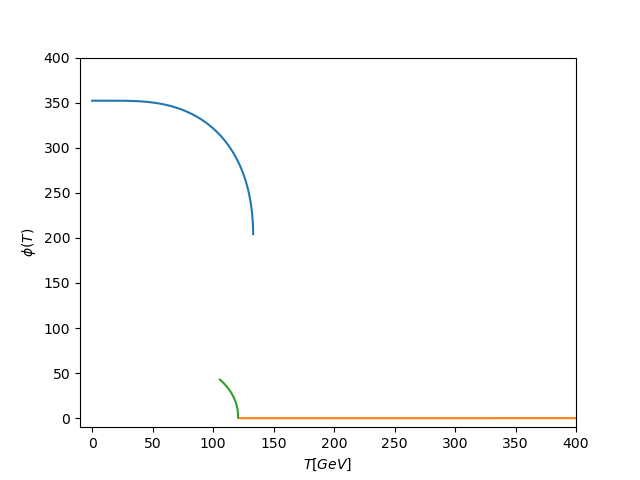}}
    \subfigure[$w=288$ GeV]{\includegraphics[width=0.3\textwidth]{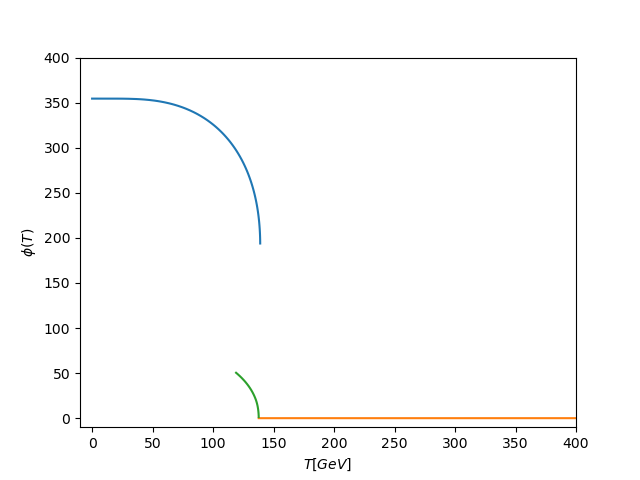}}
    \caption{Shown are strong first-order electroweak phase transitions in singlet scalar model for different scalar VEVs.}
\label{singphase}
\end{figure}

\section{Numerical Results}\label{rslts}
The independent free parameters in the model are the scalar VEV, $w$, the scalar mass $m_\text{s}$, the mixing angle $\theta$, the  DM fermions' masses $M_1$, $M_2$, and the Yukawa couplings $g_1$, $g_2$, $g_{12}$. In our scan, we restricted the scalar-Higgs mixing angle to comply the LHC14 constraint $\sin^2\theta\lesssim 0.02$ \cite{Dawson:2013bba}. We utilized the {\tt MicroMEGAs} package \cite{Alguero:2023zol}, to find the viable parameter space satisfying the observed dark matter relic density $\Omega_\text{DM} h^2\sim 0.12$ \cite{Hinshaw:2012aka, Planck:2018vyg}, and to calculate the DM-nucleon scattering cross section. We scan the parameter space over the following ranges: $1<w<500$ GeV, $1<m_\text{s}<2000$ GeV, $0<g_1,g_2<2$, $-2<g_{12}<2$, $1<M_1<1000$ GeV and $1<\delta<500$ GeV. 

Figure \ref{dm_nuc} shows the viable parameter space for DM when the observed relic density constraint is satisfied. For $m_\text{DM} < m_s$, the dominant annihilation channels are  $\chi_1 \chi_1 \to \bar f f, WW, ZZ$, and when $m_\text{DM} > m_s$, the dominant annihilation channel is  $\chi_1 \chi_1 \to$ ss. In our parameter space scan, the coannihilation contribution to the relic density is subdominant compared to the annihilation channels for the chosen mass difference $\delta$.

After applying the direct detection bounds from XENONnT \cite{XENON:2023cxc}, LZ \cite{LZ:2022lsv} and LUX \cite{LUX:2016ggv} experiments, it is evident from this figure that  DM mass heavier than around $40$ GeV evades these constraints.  The two-fermion dark matter scenario can evade direct detection bounds when $g_{12} \neq 0$, because the annihilation cross section depends on both $g_1$ and $g_{12}$, allowing an efficient annihilation cross section needed for the observed DM relic density even with small $g_1$. Since the direct detection cross section is only proportional to $g_1^2$, this decoupling lets some parameter points satisfy relic density constraints while remaining below direct detection limits (see Appendix in \cite{Ghorbani:2018hjs}).

We take only the viable parameter space below the LZ bound and  above the neutrino floor  to examine the electroweak phase transition. The numerical calculations confirms the analytic results in section \ref{sec_high}; no first-order phase transition takes place using the high-temperature approximation. To evaluate numerically the phase transition at finite-temperature described in section \ref{sec_finite}, we used the package {\tt CosmoTransitions}\cite{Wainwright:2011kj}. We examined the electroweak phase transition for all the points within the viable parameter space evading the direct detection and the DM relic density constraints depicted in Figure \ref{dm_nuc}. 

A selected number of benchmarks with different values of the scalar VEV, from $w=33$ GeV to $w=142$ GeV, are shown in Figure \ref{2stepphase}.
First, as seen in this figure, for all the benchmarks (and for the whole viable parameter space), the phase transition occurs through two steps. For all benchmarks, the phase transition starts from the symmetric phase $(0,0)$ at higher temperatures and ends to either $(v,0)$ or $(0,w)$ vacuum structure at lower temperature in the first step. In the second step, this vacuum structure changes ultimately to the vacuum structure $(v,w)$, which is required by our original assumption that both the Higgs and singlet scalar develop non-zero VEVs at zero temperature. 
Secondly, non of the phase transitions are strong in either steps.  The corresponding critical and nucleation temperatures, and the phase transition strength are given for each step in Table \ref{phasepar1}. 
In Table \ref{phasepar2}, the parameters and couplings used in the model corresponding to each phase transition diagram in Figure \ref{2stepphase} is shown. The DM mass (singlet scalar mass) ranges from $M_1=91$ GeV ($m_s=48$ GeV) to $M_1=836$ GeV ($m_s=164$ GeV). The Higgs-scalar coupling $\lambda_\text{hs}$, as well as the mixing angle $\sin\theta$, take always very small values. 

It is known that the singlet scalar model can lead to a strong first-order electroweak phase transition \cite{Ghorbani:2018yfr}. We use the {\tt CosmoTransition} package to perform phase transition calculations over the same parameter range used for the current fermion DM model. Representative benchmark points are shown in Figure~\ref{singtable}, illustrating strongly first-order phase transitions. The corresponding parameters are listed in Table~\ref{singtable}.
By comparing the singlet scalar model with the current fermion DM model, we find that the inclusion of fermions weakens the strength of the electroweak phase transition.
\begin{table}[t]
\renewcommand{\familydefault}{\rmdefault}
\centering
\setlength{\tabcolsep}{1pt} 
\renewcommand{\arraystretch}{1.2} 
\begin{tabular}{|c|c|c|c|c|c|c|c|c|}
\hline
$w$(GeV) & $\lambda_h$ & $\lambda_{hs}$ & $\lambda_s$ & $T_n$(GeV) & $T_c$(GeV) & From: $(v,w)$ & To: $(v,w)$ & $v(T_n)/T_n$ \\ 
\hline
\textbf{128.71} & 0.32 & -0.969 & 3.41 & 112.5 & 114.3 & (51.7,0) & (169.7,88.5) & 1.04 \\ \hline
\textbf{198.84} & 0.57 & -0.737 & 1.13 & 120.7 & 123.3 & (42.0,0) & (166.6,134.5) & 1.03 \\ \hline
\textbf{248.30} & 1.00 & -0.916 & 0.93 & 118.4 & 127.2 & (18.6,0) & (205.4,205.1) & 1.57 \\ \hline
\textbf{254.42} & 1.01 & -0.830 & 0.78 & 128.7 & 134.5 & (36.7,0) & (189.4,195.1) & 1.18 \\ \hline
\textbf{259.29} & 1.07 & -0.820 & 0.69 & 100.7 & 111.9 & (13.0,0) & (204.0,224.2) & 1.91 \\ \hline
\textbf{288.19} & 0.87 & -0.490 & 0.32 & 119.2 & 122.8 & (34.0,0) & (159.6,196.7) & 1.05 \\ 
\hline
\end{tabular}
\caption{Shown are the benchmarks for which the singlet scalar model gives rise to a strong first-order electroweak phase transition.}
\label{singtable}
\end{table}
\section{Conclusion}\label{cncl}
It is known that extending the Standard Model of particle physics even by a single scalar changes the type of the electroweak phase transtion (EWPT) from crossover to strongly first-order phase transition \cite{Ghorbani:2018yfr}. Consequently, adding more scalars could maintain or increase the strength of the first-order phase transition \cite{Ghorbani:2019itr}. In this work, we investigated the effect of fermion dark matter in the EWPT. To this end, we extended the singlet scalar model by two Dirac fermions, one of them being the DM candidate, to compare the phase transition behavior of the this model with that of the singlet scalar model. It is argued already that the addition of a single Dirac fermion is not enough to evade the thermal dark matter (DM) constraints \cite{Ghorbani:2018hjs}. To begin, we have presented an analytic calculation of the phase transition within the high-temperature approximation formulation. Two phase transitions $(0,0)\to(v,0)$ and $(0,w)\to(v,0)$ are not of our interest because the singlet scalar is required to gain a non-zero VEV to obtain the correct relic density for the DM fermion in the freeze-out mechanism. We have shown that no phase transition with the structures $(0,0)\to(v,w)$ or $(0,w)\to(v,w)$ occurs at high temperature. Using the full thermal corrections of the effective potential at finite temperature, and taking into account the observed DM relic density, and DM direct detection constraints, we have demonstrated that although two-step first-order phase transition is  possible in the presence of DM fermion, but these phase transitions can never be strong.  Therefore, we concluded that  the addition of fermions in the extended theories of the SM, if not changing the type, at least weaken the strength of the EWPT.

\bibliography{ref.bib}
\bibliographystyle{unsrt}

\end{document}